# Wafer-scale robust graphene electronics under industrial processing conditions

E. P. van Geest[1,2], B. Can,[1,2] M. Makurat,[1,2] C. Maheu[3,4], H. Sezen[4], M.D. Barnes[5], D. Bijl[5], M. Buscema[5], S. Shankar[5], D. J. Wehenkel[5], R. van Rijn[5], J.P. Hofmann[4], J. M. van Ruitenbeek[2]*, G. F. Schneider[1]*

[1] *Leiden Institute of Chemistry, Leiden University, Einsteinweg 55, 2333CC Leiden, The Netherlands*

[2] *Leiden Institute of Physics, Leiden University, Kamerlingh Onnes Laboratory, Niels Bohrweg 2, 2333 CA Leiden, The Netherlands*

[3] *Nantes Université, CNRS, Institut des Matériaux de Nantes Jean Rouxel, IMN, F-44000 Nantes, France.*

[4] *Surface Science Laboratory, Department of Materials- and Geosciences, Technical University of Darmstadt, Peter-Grünberg-Straße 4, 64287 Darmstadt, Germany*

[5] *Applied Nanolayers B.V. Feldmannweg 17, 2628 CD Delft, The Netherlands*

\* to whom correspondence should be addressed:

g.f.schneider@chem.leidenuniv.nl

ruitenbeek@physics.leidenuniv.nl

**Abstract**

*For commercial grade electronic devices, stable structures are required to ensure a long device life span. When such devices contain nanomaterials like graphene, it is crucial that these materials resist industrial processes and harsh environments. For environments that contain water, graphene delamination is a notorious drawback, as water intercalation and eventually liftoff readily occur in aqueous and especially in alkaline solutions. This limitation renders graphene incompatible with key wafer-processing steps in the semiconductor industry. In this work, a covalent pyrene-based adhesion layer is synthesized in a facile, two-step procedure. Through π-π interactions, the adhesion of graphene to silicon wafers was maintained under conditions that resemble harsh processes, i.e. acidic and alkaline solutions, several organic solvents, and sonication. Moreover, they could be produced with a device measurement yield up to 99.7% and reproducible device-to-device electronic performance on 4-inch silicon wafers. Our results show that a straightforward functionalization of silicon wafers with an adhesive layer can be directly applicable in industrial-scale fabrication processes, giving access to robust graphene field effect devices that are built to last long.*

**Introduction**

The lifetime of electronic devices can span over decades, as the materials used are stable and robust. Likewise, incorporating nanofabricated and two-dimensional (2D) materials should yield robust devices, too. Graphene, a 2D semimetal with unique mechanical and electrical properties,[1, 2] has been extensively explored for use in electronic devices, for example photonics and optoelectronics,[3] micro-electromechanical systems (MEMS),[4, 5] and sensors.[6-8] In practice, however, graphene on silicon wafer substrates can be easily damaged by contact with chemicals and by mechanical stress. To improve stability, annealing at high temperatures can be used to increase the adhesion energy of graphene to $SiO_2$ by 2 to 3 fold,[9, 10] and remove organic residues at the same time.[11, 12] Yet, graphene on hydrophilic surfaces such as $SiO_2$ remains prone to water intercalation between graphene and the wafer surface, and graphene delamination.[13] This is a problem, as common wafer processing steps in device fabrication require (harsh) aqueous conditions, *e.g.* acidic and alkaline conditions for etching and cleaning. The negative effect of such conditions on the graphene integrity can severely decrease the final device quality and yield and even rule out the use of these processing steps in the first place.

An attractive approach to prevent device degradation due to water intercalation and graphene delamination is to introduce a water-repelling adhesion layer on the silicon wafer. Hydrophobic groups can be introduced on the wafer surface that can create strong interactions between the graphene and the wafer while repelling water from the wafer surface. The $SiO_2$ surface can be chemically modified with a wide range of different functionalities via silane chemistry to create molecular monolayers that are highly stable and have reactive anchor groups.[14-16] Silane-based monolayers such as hexamethyldisilazane (HMDS) and octadecyltrimethoxysilane (OTS) are commonly used in electronic devices to improve their performance.[17-19] In this work, the hydrophobic pyrene group was covalently attached to the wafer surface via solution-based silane and peptide coupling chemistry with chemicals that are widely available commercially. The pyrene moiety was attached via a flexible linker to facilitate optimal alignment with the graphene sheet and promote binding. This creates a strong π-π interaction between the graphene sheet and the pyrene moieties that are covalently attached to the silicon wafer.[20, 21] We used this adhesion layer to study the stability and electronic properties of graphene field effect transistors (GFETs) that were exposed to sonication, acidic and alkaline solutions, and organic solvents that are known to be damaging to graphene on silicon wafer. The increased adhesion led to a drastic improvement of the stability of the GFETs under these harsh conditions, further widening the fabrication process window for graphene-based devices. Moreover, the procedure could be scaled to 4-inch wafers and the modified wafers could be directly integrated in large-scale fabrication of GFETs to give a very high device yield (up to nearly 100%), with similar electronic properties as devices fabricated on bare wafers.

**Results and discussion**

The surface of Si/SiO$_2$ wafers with a 285 nm thermal oxide layer were covalently functionalized with a pyrene moiety (PYRENE) using a two-step process (see Figure 1A; for the full protocol see the Supplementary Information). In short, clean oxygen-plasma-treated wafers were first chemically modified with a 5vol% solution of aminopropyl-triethoxysilane (APTES) in ethanol to introduce NH$_2$ groups at the wafer surface. Pyrene groups were then attached via a peptide coupling reaction in a triethylamine (TEA)-basified DMF solution (4 drops per 10 ml DMF) containing 1-pyrenebutyric acid (15 mM) and hexafluorophosphate azabenzotriazole tetramethyl uronium (HATU, 22 mM). After three days, the wafers were retrieved from the solution. They were then cleaned by sonication in acetone, followed by a sequential solvent rinse (acetone, water, and isopropanol (IPA)) and dried with pressurized N$_2$. Non-functionalized Si/SiO$_2$ wafers (BARE), as well as wafers that were functionalized with OTS, HMDS, APTES, or phenyltriethoxysilane (PHEN) served as controls (see Supplementary Information, Materials and Methods).

Contact angle (CA) measurements revealed that the hydrophobicity of the wafer changed after its chemical modification (Table 1 and Figure S1). Following the synthetic route, starting from a hydrophilic bare wafer, the sequential APTES- and PYRENE-functionalization increased the contact angle. The PYRENE wafers were more hydrophobic than HMDS-functionalized wafers, but less hydrophobic than OTS wafers. Atomic force microscopy (AFM) showed that the surface roughness of the wafers increased after the chemical modification as well (Table 1 and Figure S2), as indicated by the R$_a$ values. The root mean square (RMS) values on the other hand are higher than for bare wafers, due to large particles on the modified surfaces. Hence, the PYRENE modification changes the surface topology, indicative of the presence of the newly introduced functional moieties. X-ray photoelectron spectroscopy (XPS) confirmed the presence of nitrogen after the APTES reaction, while the C1s emission lines increased as well (Figure S3). For the pyrene functionalized wafers, the XPS emission lines for all the expected carbon-based bonds were identified; especially the C=C peak at 284.7 eV (which was not found in bare or APTES-functionalized wafers) demonstrated the presence of the pyrene moieties at the wafer surface (Figure 1B).

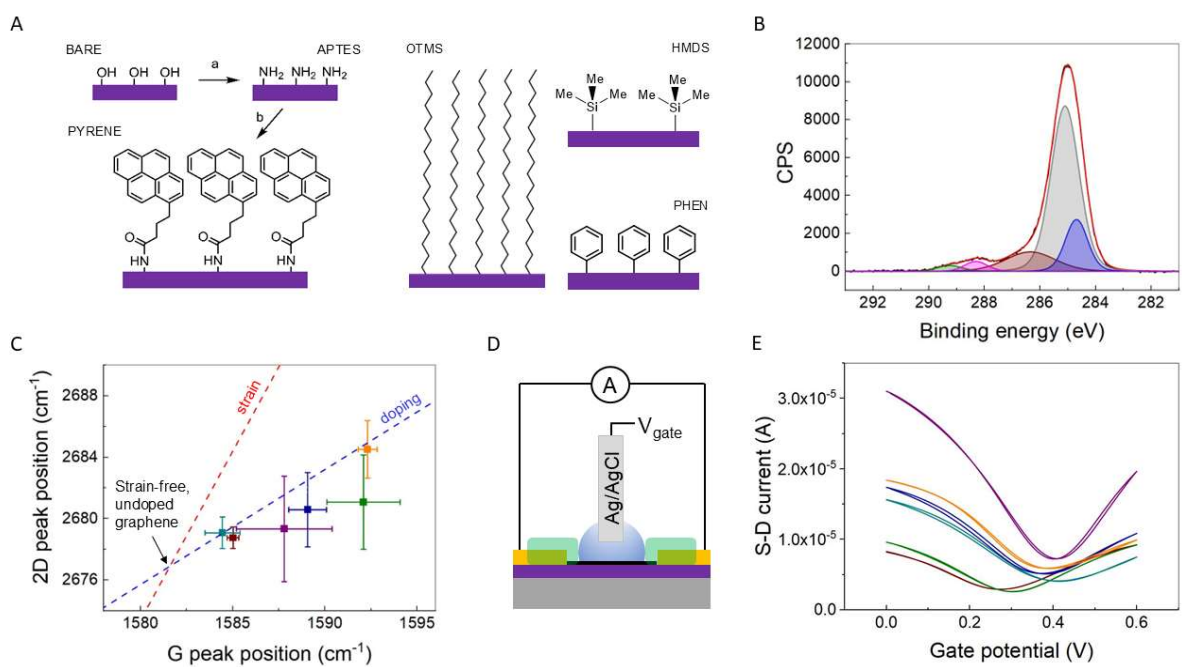

**Figure 1: Pyrene functionalization of silicon wafers and characterization of graphene on modified surfaces.**
A) Synthesis scheme for the two-step functionalization, where (a) the SiO$_2$ wafer is immersed overnight in a 5vol% solution of APTES in ethanol/water (v/v 96:4), then (b) 1-pyrenebutyric acid is reacted via a HATU peptide coupling in a TEA-basified DMF solution for three days. Both reactions are performed at room temperature. Reference surface modifications OTS, HMDS and PHEN are shown on the right-hand side of the figure panel. B) Baseline-subtracted XPS spectrum of the pyrene-functionalized wafer, deconvolution of the C1s peak (CPS = counts per second). Interpretation: 284.7 eV (blue) = sp2 C 1s, 285.0 eV (grey) = sp3 C 1s, 286.3 eV (brown) = C-N (residual amine), 288.3 eV (pink) = N-C=O (amide), 289.3 eV (green) = O-C=O (residual carboxylic group). C) Deconvolution of the Raman (average) into the strain (red) and doping (blue) components of the vector. Each data point represents four independently functionalized samples (five spectra per sample). Error bars are generated from four independently functionalized wafers. The red and blue lines indicate zero strain and zero doping axes.[22] D) Schematic representation of the liquid-gated GFET. E) Typical $I_{SD}$ vs. $V_{gate}$ plots for liquid-gated GFETs (0.1M LiClO$_4$). $V_{gate}$ was swept ten times between 0 and 0.6 V at 0.01 Vs$^{-1}$, the 10$^{th}$ cycle is displayed in the plot. Color code for panels C and E: bare (SiO$_2$) = purple, OTS = red, APTES = green, HMDS = cyan, PHEN = blue and PYRENE = orange.

**Table 1: Contact angle of modified wafers, and Raman analysis of graphene and electronic characterization of liquid-gated GFETS on the functionalized wafers**. Hys. = hysteresis. Average values are tabulated, standard deviation indicated in brackets.

| Surface | CA | AFM - roughness | | Raman - peak position | | $I_{SD}$ vs. $V_{gate}$ | |
|---|---|---|---|---|---|---|---|
| | $\theta_c$ (°) | $R_a$ (pm) | RMS (pm) | G (cm$^{-1}$) | 2D (cm$^{-1}$) | $V_{CNP}$ (mV) | Hys. (mV) |
| Bare | 35 (7) | 157 (26) | 266 (66) | 1587.8 (2.6) | 2679.3 (3.4) | 379 (86) | 14 (44) |
| OTS | 92 (5) | 240 (55) | 529 (215) | 1585.0 (0.3) | 2678.7 (0.7) | 282 (30) | 12 (15) |
| APTES | 49 (7) | 203 (62) | 732 (665) | 1592.1 (2.0) | 2681.1 (3.1) | 303 (55) | 7 (12) |
| HMDS | 61 (7) | 194 (32) | 396 (80) | 1584.4 (1.0) | 2679.1 (1.0) | 377 (36) | 10 (12) |
| PHEN | 42 (2) | 305 (206) | 688 (577) | 1589.1 (1.0) | 2680.6 (2.4) | 353 (63) | 2 (9) |
| PYRENE | 69 (5) | 231 (66) | 565 (298) | 1592.3 (0.5) | 2684.5 (1.9) | 379 (50) | -8 (7) |

To study how the different surface modifications influenced the physical properties of the transferred graphene layer, both Raman and liquid gating experiments were conducted. Graphene (grown on a Cu film by chemical vapor deposition) was transferred to a (modified) wafer using PMMA-assisted transfer. Raman analysis of the graphene sheets ($\lambda_{ex}$ = 532 nm, P = 2.30 mW) was performed after PMMA was removed. On all graphene samples, the common peaks indicative of the presence of graphene were found (*i.e.* the 2D, G, and occasionally the damage-indicating D peak, see Figure S4).[23] The 2D peak was plotted against the G peak position (Figure 1C), which gives information about the strain and doping of graphene.[22] For each consecutive step of the functionalization strategy (BARE-APTES-PYRENE), the doping of graphene increases (as the data points moves further right along the blue doping line), and tensile strain is increased due to the APTES layer. However, the graphene is not strained - as the (2D, G) point is exactly on the blue (doping) line – when it is on the pyrene layer (which is also the case for OTS and HMDS); it appears that hydrophobic substrates reduce the strain in the transferred graphene layer.

The effect of the different surface modifications on the electronic properties of liquid-gated GFETs was less obvious. GFETs were fabricated on the modified wafers (Supplementary Information); through protecting the Au electrodes with a protective resin, the fabricated devices were made suitable for liquid gating (Figure 1D). In such gating experiments, the presence of p-type dopants (*e.g.* intercalated water, polymer residues, or charged surface traps) near the graphene is reflected by a positive shift of the voltage at the charge neutrality point ($V_{CNP}$) of graphene. Molecular layers can reduce such doping effects, for example by screening the graphene from the dopant shifting the $V_{CNP}$ towards neutrality.[24-27] They can also reduce the presence of dopants in the first place, as hydrophobic layers reduce water adsorption for example. The reduction of doping was indeed most prominent for the highly hydrophobic layer OTS, which displayed the lowest $V_{CNP}$ values. For the APTES-layer $V_{CNP}$ was relatively low too; this is in line with reported n-doping by the electron-donating $NH_2$ groups.[28] However, the $V_{CNP}$ for the GFETs on PYRENE wafers (Figure 1E and Table 1) was the same as for the unmodified hydrophilic bare surfaces and similar $V_{CNP}$ values were found for GFETs on the HMDS and PHEN functionalized wafers. It thus seems that the pyrene layer did not have a measurable effect on the doping of graphene. This appears to be in contradiction with the Raman analysis regarding doping in Figure 1C, yet it should be considered that the presence of an electrolyte during liquid gating makes that the environment of the graphene was completely different for the different experiments.

Another series of GFET devices was prepared to study the mechanical stability and electronic properties of graphene under mechanical stress induced by sonication in acetone via liquid gating. Here, sonication was done stepwise to monitor the electronic properties of the devices as the sonication time increased. The devices on APTES and PYRENE wafers showed the highest stability towards sonication since all of them remained intact. On the other hand, devices on other surfaces suffered more than a 25% yield reduction due the loss of electrical contact between the gold electrodes (Figure 2A and Table 2). At the same time, for GFETs on APTES and pyrene-modified surfaces the resistance at $V_{CNP}$ ($R_{max}$) did not increase significantly, while for all other surface types $R_{max}$ increased already after a relatively short sonication period (<5 min, Figure S5).

Analyzing $V_{CNP}$ particularly, we observe that sonication also affects the electronic properties of the GFETs (Figure 2B). For all devices, $V_{CNP}$ increased in the first few minutes of sonication. After this initial increase, the behavior of surviving devices started to vary between the different surface modifications. While $V_{CNP}$ drops for increasing sonication times for OTS as well as for BARE, it stabilized for the other surface modifications after 10 minutes of sonication to values that are typically higher than their initial $V_{CNP}$. We also analysed the hysteresis ($V_{CNP, forward} - V_{CNP, backward}$) of the GFETs. Hysteretic behavior can occur due to charge transfer or trapping (*e.g.* at the graphene/oxide interface or at defect sites in graphene), or due to the slow rearrangement of polar species near the graphene, which can be for example impurities, but also physisorbed or intercalated water.[29-31] Decreased hysteresis may therefore indicate weakened interactions of the GFET with (physisorbed and/or intercalated) water and thus the effectiveness of the wafer coating in decreasing the adverse effects of water on graphene devices. In our case, devices on BARE, OTS, and HMDS-modified wafers showed hysteresis directly after fabrication, and this increased rapidly as they underwent sonication (Figure 2C). This increase could arise from progressive water intercalation and/or the formation of damages in graphene when it is weakly bound to the surface. For APTES, the initially increased hysteretic behavior decreases slowly as sonication progresses, which may indicate that polar species (*e.g.* impurities) are being removed from the graphene surface during sonication. Interestingly, PHEN- and pyrene-modified devices were practically hysteresis-free throughout the entire experiment (although one PHEN-device did not survive the sonication process). The measured stability with respect to sonication indicates that the PYRENE layer performed best in protecting the graphene from water intercalation and mechanical damage.

Lastly, we calculated the hole and electron mobility ($\mu_{hole}$, $\mu_{electron}$) from the slope of the liquid gating curves (Figure 2D; for $\mu_{electron}$, see Figure S6). In most cases, $\mu_{hole}$ and $\mu_{electron}$ decreased for increasing sonication times. However, for APTES the mobility appears to be not affected, and for PYRENE it even increases. Sonication thus has a strong effect on the carrier mobility of graphene, and surface modifications can modulate its efficacy. Overall, the device survival, unchanged sheet resistance, $V_{CNP}$ stabilization, hysteresis-free behavior and increasing carrier mobility of the device on the pyrene-modified wafers lead us to conclude that incorporation of the pyrene layer not only increased the robustness of GFETs under sonication as it was reported before[21] but preserves the electronic properties of the devices as well.

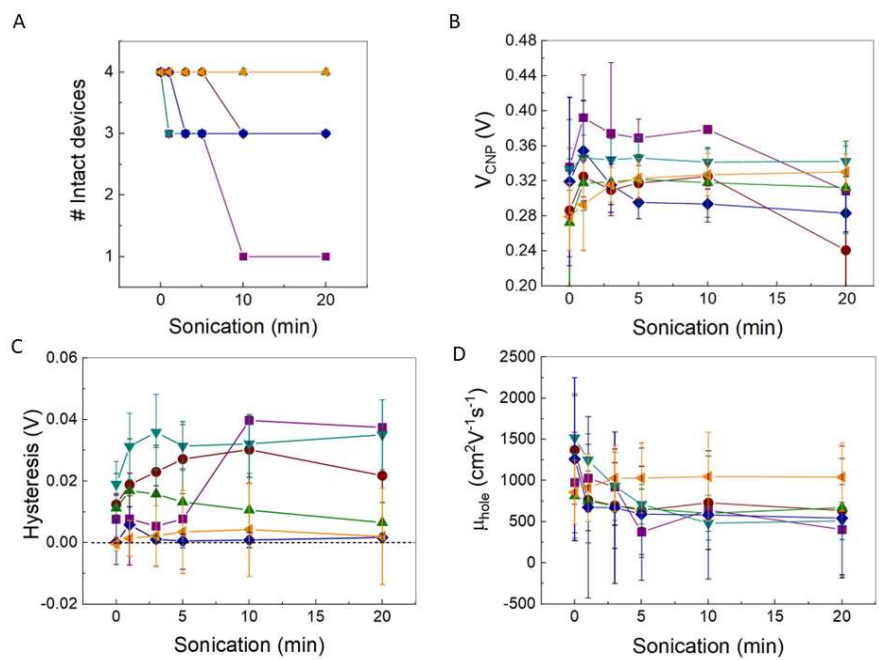

**Figure 2: Stability of GFETs during sonication in acetone, analysed by liquid gating.** A) Device survival during sonication. # Intact devices indicates the number of devices that remained intact after sonication. B) $V_{CNP, forward}$ evolution during increased sonication times. C) Device hysteresis ($V_{CNP,forward} - V_{CNP,backward}$) evolution as a result of sonication. D) Evolution of hole mobility $\mu_{hole}$ during sonication. Liquid gating was performed using 0.1M LiClO$_4$ as the electrolyte, $V_{gate}$ was swept ten times between 0 and 0.6V at 0.01 Vs$^{-1}$, the data of the last cycle is reported in the plots. Color coding for all panels: bare (SiO$_2$) = purple, OTS = red, APTES = green, HMDS = cyan, PHEN = blue and PYRENE = orange.

**Table 2: Liquid gating ($I$ vs. $V_{gate}$) results for GFETs constructed on surface modified wafers, and for GFETs after sonication (t = 20 min)**. $dI/dV_{t=20min}$ values are reported relative to $\Delta dI/dV_{t=0min}$. Different sets of GFETs were used for the graphene characterization and sonication experiments. Average values are tabulated, standard deviation indicated in brackets.

| | | | | $I_{SD}$ vs. $V_{gate}$ – sonication | | | | | |
|---|---|---|---|---|---|---|---|---|---|
| | Survival | | $\Delta R_{max}$ | $V_{CNP}$ (mV) | | Hysteresis (mV) | | $\mu_{hole}$ (x10$^3$ cm$^2$V$^{-1}$s$^{-1}$) | |
| **Surface** | $n_{t=0}$ | $n_{t=20}$ | $R_{t=20} / R_{t=0}$ | 0 min | 20 min | 0 min | 20 min | 0 min | 20 min |
| Bare | 4 | 1 | 2.38 (-) | 336 (54) | 309 (-) | 8 (8) | 37 (-) | 0.98 (0.61) | 0.41 (-) |
| OTS | 4 | 3 | 3.80 (2.6) | 286 (53) | 241 (84) | 12 (3) | 22 (15) | 1.37 (0.66) | 0.64 (0.78) |
| APTES | 4 | 4 | 1.19 (0.4) | 272 (73) | 312 (53) | 11 (11) | 7 (7) | 0.82 (0.52) | 0.67 (0.28) |
| HMDS | 4 | 3 | 2.66 (1.2) | 333 (24) | 342 (17) | 19 (7) | 35 (11) | 1.52 (0.53) | 0.51 (0.23) |
| PHEN | 4 | 3 | 3.03 (3.8) | 319 (96) | 283 (21) | 0 (7) | 2 (1) | 1.26 (0.99) | 0.54 (0.72) |
| PYRENE | 4 | 4 | 1.01 (0.2) | 279 (38) | 330 (20) | 0 (2) | 2 (16) | 0.86 (0.39) | 1.04 (0.41) |

To examine if the wafer functionalization would perform similarly at industrial scale, intact 4-inch wafers were chemically modified with the pyrene layer and coated with graphene via dry transfer, using an industry-integrated process. For comparison, graphene was transferred on bare and OTS-modified wafers as well. Contact angle and AFM inspection of the wafers prior to transfer showed that surfaces were successfully modified (see Table S1). After transfer, the graphene layers were inspected with optical microscopy and Raman spectroscopy mapping (Figure S7-9 and Table S1 and S2). Remarkably, the effect of the surface modification was directly visible. For bare and pyrene-coated wafers, the graphene was intact at the surface (99% coverage for both), for OTS the sheet was badly damaged, *i.e.* ripped and in parts not transferred at all (coverage = 94%). A wafer-scale Raman analysis revealed that graphene on bare wafer on average is clearly tensile-strained and doped. On the other hand, OTS and PYRENE wafers show negligible strain and strong reduction in doping (Figure 3A and Table 3). Notably, the spread of the data points in the direction of the doping vector component (Figure S10) was small for graphene on PYRENE wafers as compared to bare wafers. For OTS wafers the wide data point spread suggests non-homogeneous graphene layer that has *e.g.* folds, ripples, and damages. The pyrene-coated wafer thus enabled low-strain and charge-screened graphene with high coverage and quality on wafer scale. We note that these results, most notably the doping of graphene on pyrene-coated wafers, are different from the Raman analysis we have performed at smaller scale. We ascribe this difference to the different transfer methods and source materials.

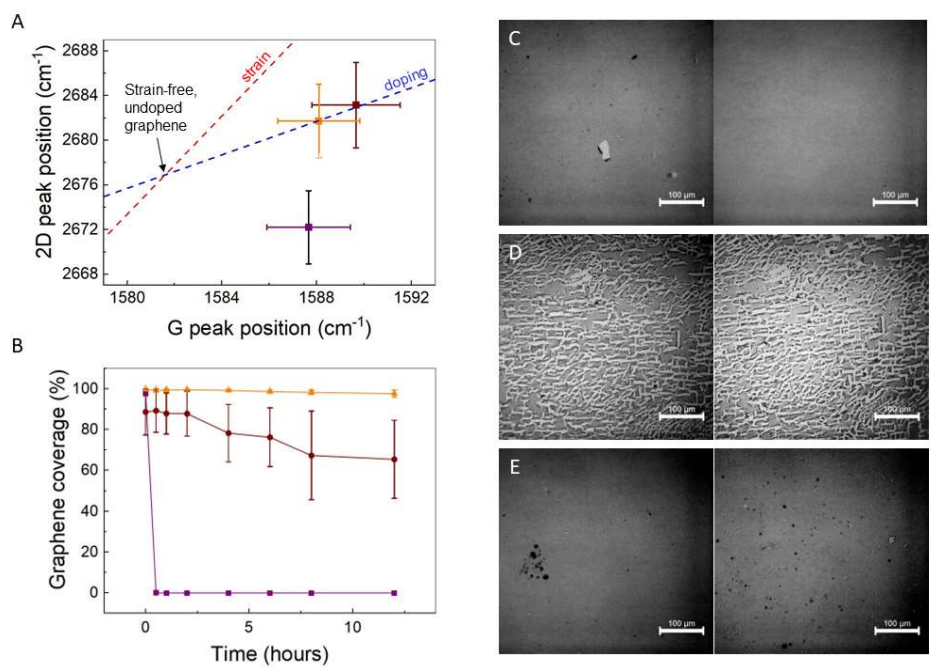

**Figure 3: Full wafer (4") scale dry transfer of graphene on modified wafers.** A) Raman analysis of graphene on functionalized wafers (bare = purple, OTS = red, PYRENE = orange). Red and blue diagonal lines represent the axes for strain and doping. Deconvolution of the Raman (average) into the strain (red) and doping (blue) components of the vector. B) Graphene coverage vs. time for samples immersed in KOH (0.5M in water, room temperature), for graphene on BARE, OTS and PYRENE wafers (purple, red, and orange respectively). Samples were optically inspected (three samples for each functionalization, four photos per sample) and delamination was determined as the percentage of graphene coverage in the centre of the image (a circle with ⌀ = 150 μm). C-E) Contrast-enhanced optical images (magnification x20) for graphene on bare (C), OTS (D), and PYRENE (E) wafers immersed in KOH for 0 and 12h (left and right, respectively). The tears in the right image of (E) indicate that graphene has small damages but has not delaminated.

**Table 3: Full wafer-scale graphene on modified surfaces - characterization.** Graphene coverage values tabulated for 12h in 0.5M KOH, 12h in glacial acetic acid (AA), and 18h in pure NMP. Average values are tabulated, standard deviation indicated in brackets.

| Surface | Raman - peak position | | Graphene coverage | | |
|---|---|---|---|---|---|
| | G (cm$^{-1}$) | 2D (cm$^{-1}$) | KOH (%) | AA (%) | NMP (%) |
| Bare | 1587.7 (1.8) | 2672.2 (3.3) | 1 (0) | 30 (41) | 3 (3) |
| OTS | 1590.0 (1.9) | 2683.2 (3.8) | 65 (19) | 82 (16) | 22 (18) |
| PYRENE | 1588.1 (1.7) | 2681.7 (3.3) | 98 (2) | 99 (0) | 99 (1) |

We also evaluated if the graphene on modified wafers could survive in solutions that are commonly used in industrial chip processing. After dicing into 1x1 cm² chips, these chips were immersed in either 0.5M KOH, glacial acetic acid (AA) or N-methyl pyrrolidone (NMP), solvents that are commonly used for silicon etching and cleaning. It was evident that the surface modifications helped to preserve the graphene quality, as the graphene delaminated much slower (OTS) or hardly at all (PYRENE) as compared to the bare wafers. For example, while graphene on bare wafers is completely removed after 30 minutes in the KOH solution, on PYRENE wafers the graphene was only slightly damaged after being immersed for 12 hours (98% coverage) (Figure 3B-E and Table 3). Similarly for acetic acid, the graphene on bare wafer was mostly removed after 4 hours, while the graphene coverage on OTS and PYRENE wafers did not change (Figure S11). When immersed overnight in NMP, the graphene was only found to be stable on the pyrene-coated wafers; on bare it was completely removed, while for OTS the graphene had mostly rolled up on itself, forming scrolls (Figure S12). Overall, the tested surface modifications help to preserve the graphene quality under these chemical conditions, and PYRENE shows the best performance.

Finally, we have fabricated large arrays of GFETs (2-terminal and 4-terminal) on intact BARE and pyrene-coated wafers (4-inch, $SiO_2$ thickness = 285 and 90 nm, one of each) to study the influence of the adhesive layer when incorporated in the industrial processes for graphene electronics. The OTS functionalization was excluded due to the initial large-scale mechanical damage to the graphene layer after transfer observed in the solvent delamination tests, preventing a good device yield. Finished devices were characterized via back-gating using an automatic probe station. A typical result is displayed in Figure 4A. Statistical analysis was performed on a multitude of devices (Table 4). It should be noted that the results for BARE devices (see also Figure S14 and S15) were recorded around the same time of the year as the PYRENE wafers were measured, but the devices were not necessarily fabricated on wafers of the same wafer batch and the gate-capacitance profile as a function of gate voltage bias can vary significantly between wafer batches. Therefore, the results serve as a guideline, not a one-to-one comparison.

The first performance parameter we considered is the device measurement yield, which is defined as the ratio of devices successfully measured (with $V_{CNP}$ within the sweep range) and the total number of devices of a measurement run. Importantly, this definition excludes devices which are intact and showing typical graphene behavior but with their $V_{CNP}$ out of the measurement range. The yield is thus not solely based on functional, closed-circuit devices. The PYRENE wafers showed a high measurement yield of 86.2% and 99.7%, for the wafers with 285 nm and 90 nm $SiO_2$ thickness respectively. These yields were significantly higher than the ones for BARE (31.4% and 89.0% for 285 and 90 nm oxide). These results imply that the adhesive layer helps to increase the device yield, likely by improving the mechanical stability of the graphene during the fabrication process by increasing adhesion to the substrate.

We further analysed the electrical properties of the devices (4-terminal measurements) on the pyrene-coated wafers and found that they were similar to GFETs on BARE wafers. Devices fabricated on the wafer with 285 nm $SiO_2$ thickness provided an average $V_{CNP}$ of 75 V with a hysteresis of a few volts, and hole mobility of 1990 $cm^2V^{-1}s^{-1}$ (see Figure 4B and C). The electron mobility was lower, averaging at 943 $cm^2V^{-1}s^{-1}$, picturing an asymmetric gate sweep profile (Figure 4D). When GFETs were produced on a wafer with 90 nm $SiO_2$, the electronic characteristics were similar to the ones on 285 nm $SiO_2$ (Table 4 and Figure S13), although the average $V_{CNP}$ is lower at 21 V; this is to be expected as gating is less efficient with thicker oxide, and the value found is normal for GFETs on this type of wafers. Hysteresis however was significantly higher for these devices than for the 285 nm wafer. With respect to the BARE wafers, the electronic properties appear to be preserved as hole and electron mobilities are similar or slightly higher on the PYRENE wafer for both 90 and 285 nm oxide layer (although the hysteresis was smaller on the BARE wafers of both thicknesses). These results confirm that the PYRENE layer can be applied directly on full wafers to obtain high device yields and preserving the important device parameters.

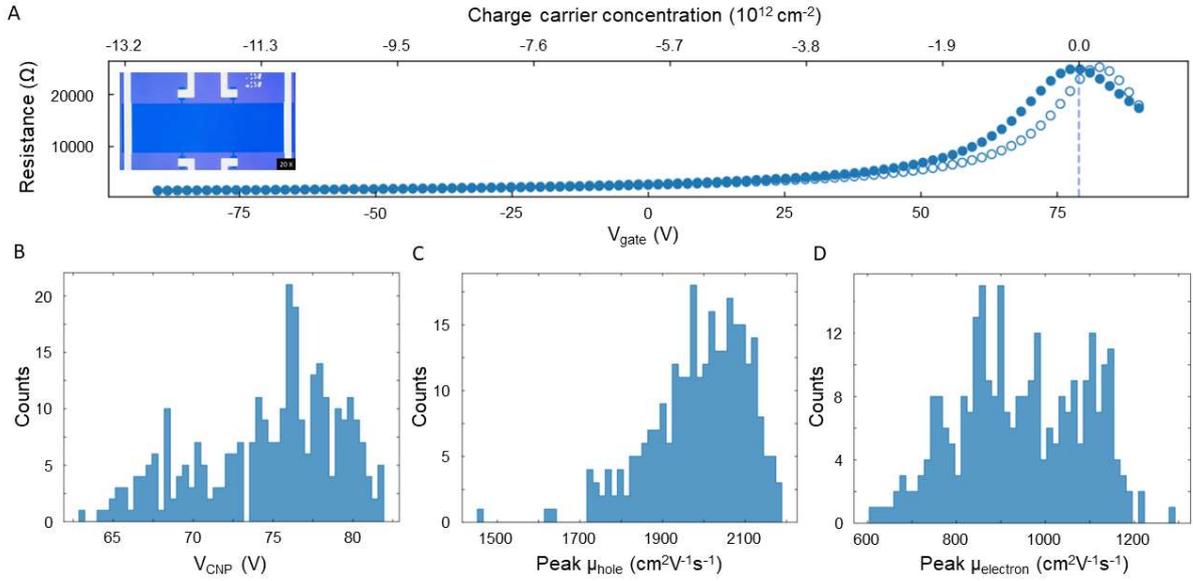

**Figure 4: GFET characterization (4-terminal) of wafer-scale device fabrication.** A) Typical gate sweep for a GFET on a PYRENE wafer with 285 nm $SiO_2$ layer. Forward (backward) sweep in filled (empty) circles. B-D): $V_{CNP}$, $\mu_{hole}$, and $\mu_{electron}$ histograms for devices fabricated on a pyrene-coated $Si/SiO_2$ wafer with 285 nm $SiO_2$ thickness. Gate sweeps were performed at 10 V/s. Numerical values are presented in Table 4.

**Table 4:** Statistical analysis for GFETS on BARE and PYRENE wafers. Average values are tabulated, standard deviation indicated in brackets.

| Surface + oxide | Device fabrication | | Charge neutrality point (V) | | Carrier density (x$10^{12}$ cm$^{-2}$) | | Transconductance Mobility (x$10^3$ cm$^2$V$^{-1}$s$^{-1}$) | | | |
|---|---|---|---|---|---|---|---|---|---|---|
| | | | | | | | 2-terminal | | 4-terminal | |
| | Count | Yield | $V_{CNP,up}$ | $V_{hyst}$ | $n_0$ | $n_{hyst}$ | $\mu_{hole}$ | $\mu_{electron}$ | $\mu_{hole}$ | $\mu_{electron}$ |
| BARE 285 nm | 830 | 31.4 | 79.3 (3.0) | -1.7 (0.7) | 6.00 (0.23) | -0.13 (0.06) | 1.76 (0.08) | 0.73 (0.31) | 1.74 (0.11) | 0.671 (0.27) |
| BARE 90 nm | 833 | 89.0 | 28.7 (3.0) | -2.5 (0.8) | 6.85 (0.71) | -0.59 (0.19) | 1.53 (0.15) | 1.08 (0.17) | 1.58 (0.16) | 1.10 (0.14) |
| PYRENE 285 nm | 325 | 86.2 | 74.8 (4.4) | -5.0 (1.4) | 5.66 (0.33) | -0.38 (0.1) | 2.10 (0.09) | 1.01 (0.15) | 1.99 (0.12) | 0.94 (0.14) |
| PYRENE 90 nm | 586 | 99.7 | 21.0 (2.5) | -12.8 (1.0) | 5.02 (0.58) | -3.06 (0.17) | 1.63 (0.05) | 0.76 (0.10) | 1.59 (0.08) | 0.72 (0.10) |

## Conclusions

Silicon wafers were successfully functionalized with a pyrene-based graphene adhesion layer, in a facile two-step procedure. The resulting increased hydrophobicity upon pyrene functionalization is reflected in an increase in the contact angle, and the introduction of nitrogen and sp$^2$ carbon is confirmed by XPS. This confirms the successful surface modification with the pyrene moieties. The incorporation of this surface layer ensured the graphene devices fabricated on top of the modified wafers showed preserved electronic properties and made them more resistant to mechanical damage. Delamination of graphene in acidic and basic solutions was strongly reduced, showing that the strong interactions between the pyrene molecules and the graphene prevent that graphene is lifted off under these conditions. Finally, the advantage of the pyrene layer was demonstrated on a full-wafer scale, as device yield of close to 100% was achieved, maintaining similar electronic properties as on standard, bare wafers. This ultimately shows that modified wafers with the adhesive layer can be directly integrated in graphene device fabrication on the industrial scale. These will benefit from the increased stability under harsh conditions which promotes further opportunities, for example in sensors where graphene needs to be further functionalized and regenerated.

**Conflict of interest**

The authors declare that a patent application has been filed on this work.

**Acknowledgements**

The work was supported by the Nederlandse Organisatie voor Wetenschappelijk Onderzoek (NWO project no.16671, OTP-TTW). The authors thank all the partners within the user committee, particularly Applied Nanolayers, Leiden University Medical Center, Leiden Probe Microscopy, Base Clear, Future Genomics Technologies and GenomeScan) for their support and contributions during discussions. DB, MB, SS, RR and DW acknowledged funding from the European Union's Horizon 2020 research and innovation programme under Grant Agreement SPRING No. 881273. N.V.V Bluemel is thanked for contributions in scientific discussions.

Supplementary Information

**Materials and methods**

Chemicals were purchased at Sigma Aldrich or Brunschwig Chemie and used without further purification unless stated otherwise. Monolayer graphene on copper was purchased from Graphenea. Graphene on copper was spin-coated with poly(methyl methacrylate) (PMMA, 6% in anisole, Allresist GmbH, AR-P 662.06; 4000 rpm for 60 s), heated at 85°C for 10 minutes, then back-etched in oxygen plasma (0.30 mbar, 100 W, 2 minutes) and transferred on a cleaned wafer (sonication in acetone for 5 min, then rinsed with acetone, MilliQ and isopropanol (IPA), then treated with $O_2$ plasma (0.30 mbar, 100 W, 2 minutes).

Wafer surface functionalization was performed using silane chemistry. First, bare wafer pieces or an intact 4-inch wafer were cleaned by sonication in acetone for 5 minutes, then rinsed with acetone, ultra-pure water and IPA, then blown dry with pressurized nitrogen. Next, the wafer was treated with $O_2$ plasma (0.30 mbar, 100 W, 2 minutes), and directly transferred into a solution of 5vol% of the corresponding silane in 96% ethanol in the case of aminopropyl-triethoxysilane (APTES), hexamethyldisilazane (HMDS), and phenyltriethoxysilane (PHEN); for octadecyltrimethoxysilane (OTS) a 5vol% solution of the silane in hexane was used. Wafers were kept in solution to react overnight at room temperature. Then, they were taken out of the silane solution and sonicated in acetone for 5 minutes, rinsed with acetone, ultra-pure water and IPA, and blown dry with pressurized nitrogen. The BARE wafers used as reference underwent the same procedure, except that the wafers were immersed in 96% ethanol without any additives.

To produce PYRENE-coated wafers, clean APTES-modified wafer pieces were immersed in a solution of hexafluorophosphate azabenzotriazole tetramethyl uronium (HATU, 22 mM, 1.5 eq, 85 mg) and 1-pyrenebutyric acid (15 mM, 1.0 eq, 43 mg) in 10 ml DMF that was basified using 4 drops of triethylamine (TEA). For an intact 4-inch wafer, all amounts were multiplied by a factor 5. The wafers were kept in solution to react for three days at room temperature. Afterwards, the wafers were removed from solution, sonicated in acetone for 5 minutes, rinsed with acetone, ultra-pure water and IPA, and blown dry with pressurized nitrogen.

Graphene was transferred using a PMMA-assisted transfer method by etching copper with an ammonium persulfate solution (0.2 M in ultra-pure water) and rinsing the PMMA-graphene film by transferring it in three MilliQ baths consecutively, after which this film was transferred on the wafer by bottom fishing the floating film from below. Next, water was allowed to gently evaporate at 45°C; when water was evaporated, the coated wafer was heated at 150°C for 15 minutes. After successful transfer, the PMMA layer was removed by immersing the wafer in acetone for 10 minutes, then rinsing gently with acetone, ethanol and isopropyl alcohol, and blowing the wafer dry with pressurized nitrogen.

For the industrial graphene experiments, CVD graphene was dry-transferred on intact, modified 4-inch Si/SiO$_2$ wafers (Prime grade, p-doped, 90 or 285 nm SiO$_2$ (dry), single-side polished, Siegert Wafer GmbH), which were cleaned by sonication in acetone, DI water, and isopropanol (5 minutes in each solvent) and blown dry with pressurized nitrogen.

Oxygen plasma was generated using a capacitively coupled plasma system with radio-frequency of 40 kHz and 200 W power from Diener electronic (Femto), employed at room temperature. Spin coating was performed using a POLOS SPIN150i tabletop spin coater.

**Characterization**

Optical images were obtained using a Leica DM2700 M Brightfield microscope fitted with Leica MC120 HD camera. Contact angles measurements were performed using a Ramé-Hart 250 goniometer (Netcong, NJ) in combination with the DROPimage advanced v 2.8 software. AFM images were recorded on a JPK Nanowizard Ultra in intermittent contact mode at room temperature in air, using Olympus micro cantilevers (OMCL-A160TS-R3) with a nominal resonance frequency of 300 kHz. Raman spectra were recorded on a Witec Alpha500 R Raman spectrometer using a 532 nm laser at low power (0.23 mW) and a 100x objective (lateral resolution 200-300 nm). ThermoFisher's Escalab-250 was used for XPS measurements. A monochromic Al Kα X-ray source (hv = 1486.68 eV) with a 650 µm spot size was employed. The XPS measurement chamber has a higher vacuum level than $5 \times 10^{-10}$ mbar. The high-resolution core-levels and survey spectra were obtained at pass energies of 20 eV and 50 eV, respectively. All XP spectra were calibrated with C 1s set to 285.0 eV for adventitious carbon. Electrical characterization of devices was using two Keithley Sourcemeters - model 2450 (one for the electrical measurement and one to supply a gate voltage) and Kickstart V2 software.

**Supplementary Figures and Tables**

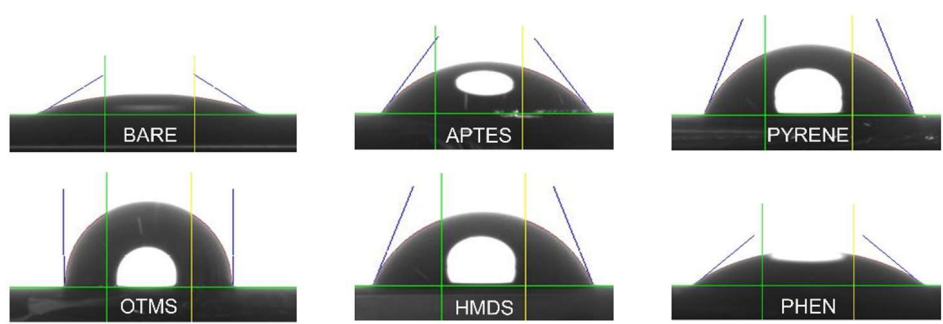

Figure S5: Optical images of CA measurements. Blue lines indicate contact angle measurement.

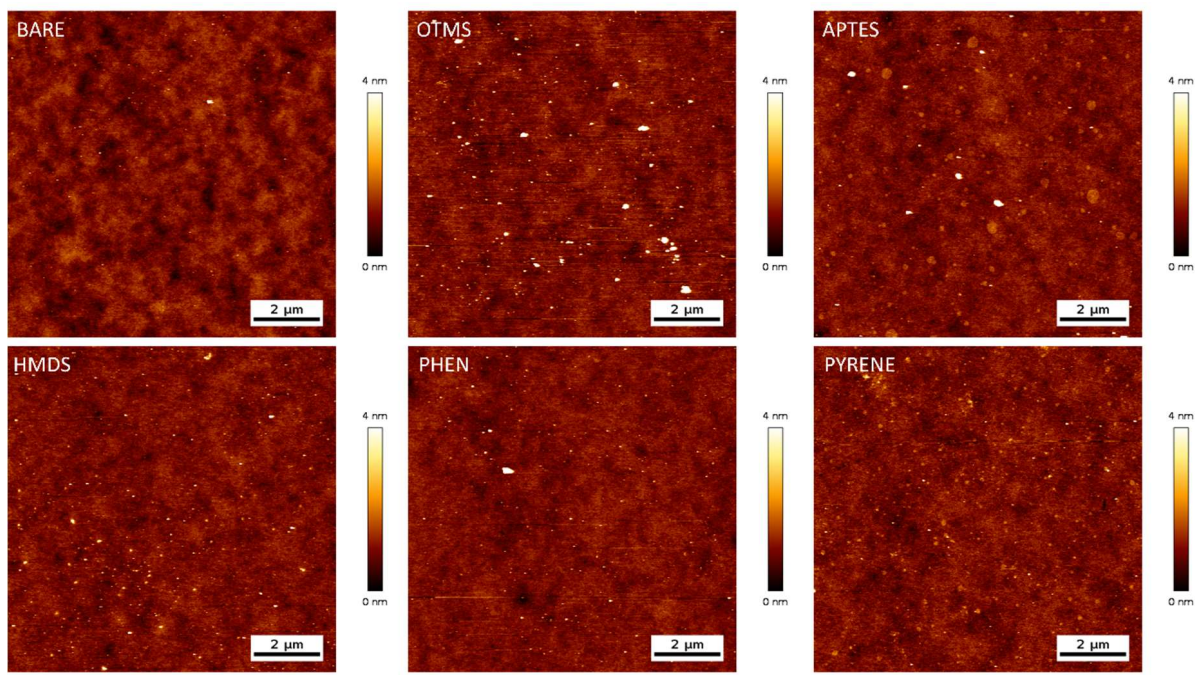

**Figure S6:** AFM images of chemically modified wafers.

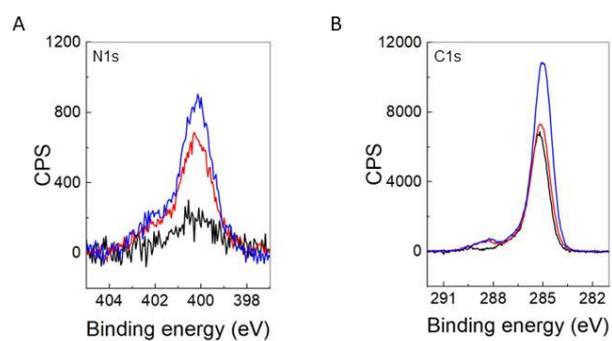

**Figure S7: Baseline-subtracted XPS spectra of bare, APTES, and pyrene-functionalized wafers. N 1s (A) and C 1s (B) core level spectra for bare, APTES-, and pyrene functionalized wafer (black, red and blue respectively). CPS = counts per second.**

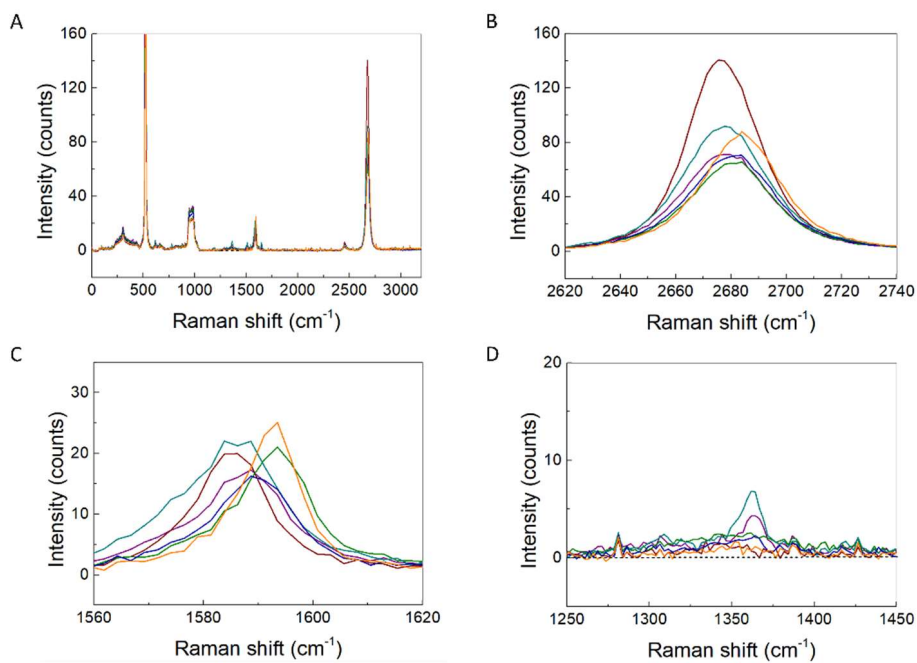

**Figure S8: Raman analysis of graphene on chemically modified wafers. A) Full spectrum. B) Zoom on 2D peak region. C) Zoom on G peak region. D) Zoom on D peak region. Colour code: bare (SiO$_2$) = purple, OTMS = red, APTES = green, HMDS = cyan, PHEN = blue and PYRENE = orange.**

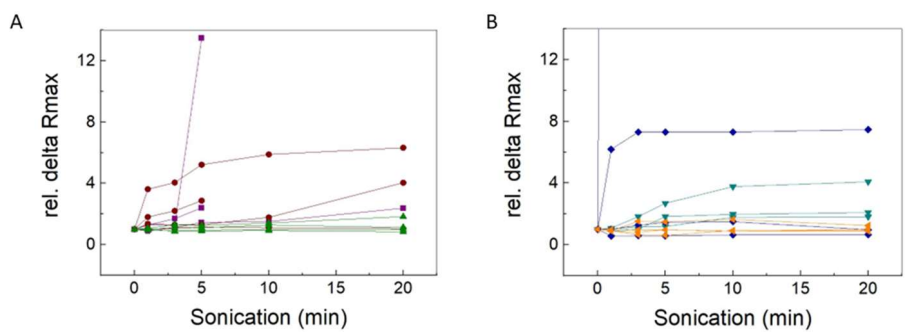

**Figure S9: Individual values relative resistance variation ($R_{max,t}$ / $R_{max,0}$) for devices that were sonicated in acetone. Panel A displays BARE, OTMS and APTES, panel B displays HMDS, PHEN and PYRENE. Colour code: bare (SiO$_2$) = purple, OTMS = red, APTES = green, HMDS = cyan, PHEN = blue and PYRENE = orange.**

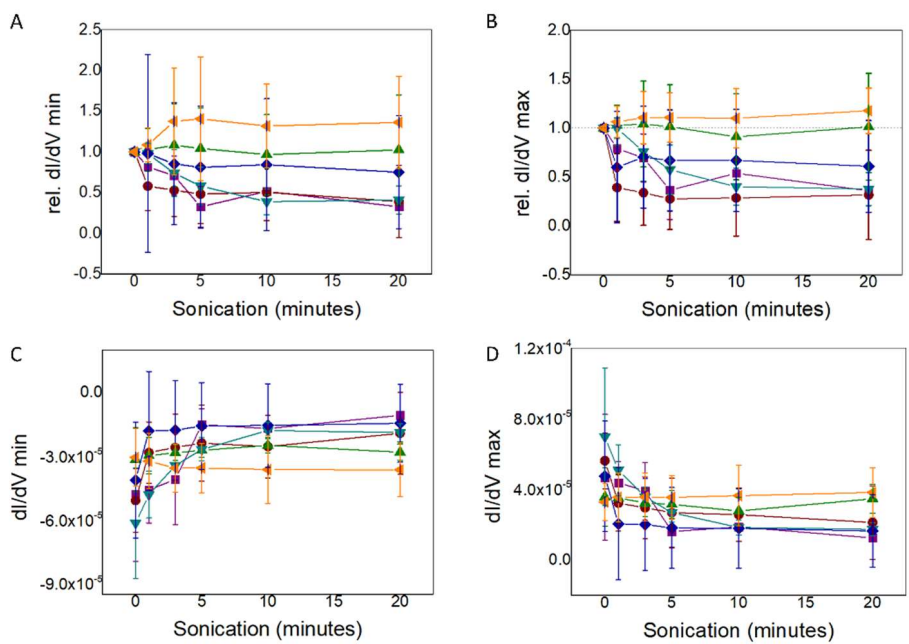

**Figure S10:** Evolution of the minimum and maximum of $dI/dV_{gate}$ (absolute and relative to $dI/dV_{gate}$ at t = 0 min) during sonication. A) relative minimum, B) relative maximum, C) absolute minimum, and D) absolute maximum of $dI/dV_{gate}$. Liquid gating performed using 0.1M LiClO$_4$, $V_{gate}$ was swept 10 times between 0 and 0.6V at 0.01Vs$^{-1}$, the data of the last cycle was used to construct the plots. Colour coding for all panels: bare (SiO$_2$) = purple, OTMS = red, APTES = green, HMDS = cyan, PHEN = blue and PYRENE = orange.

**Table S5:** Surface characterization of modified 4-inch wafers for industrial graphene application. Average values are tabulated, standard deviation indicated in brackets.

| Surface | CA | AFM before transfer | | AFM after transfer | | Raman - peak position | |
|---|---|---|---|---|---|---|---|
| | $\theta_c$ (°) | Ra (pm) | RMS (pm) | Ra (pm) | RMS (pm) | G (cm$^{-1}$) | 2D (cm$^{-1}$) |
| Bare | 42 (-) | 175 (-) | 220 (-) | 425 (78) | 1104 (221) | 1587.7 (1.8) | 2672.2 (3.3) |
| OTS | 91 (5) | 166 (-) | 384 (-) | 1878 (335) | 3674 (744) | 1590.0 (1.9) | 2683.2 (3.8) |
| PYRENE | 69 (3) | 251 (-) | 659 (-) | 894 (231) | 2200 (633) | 1588.1 (1.7) | 2681.7 (3.3) |

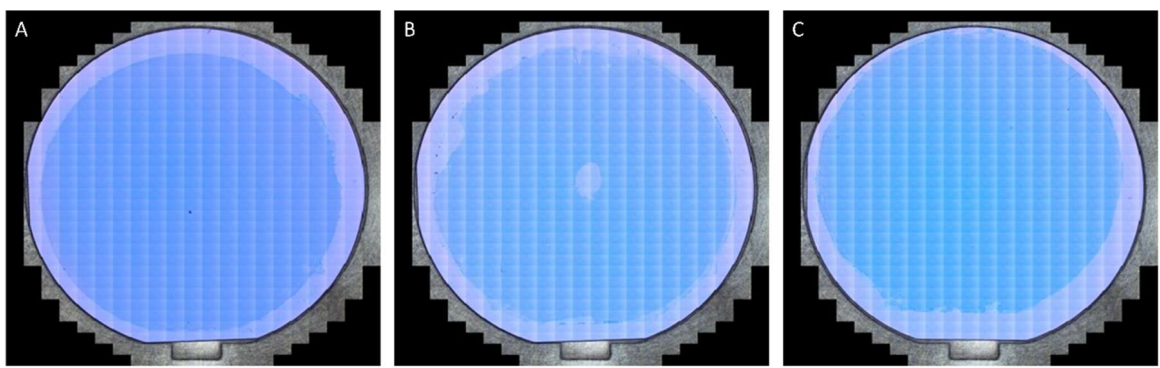

**Figure S11:** Dry-transferred graphene on 4-inch modified wafers. Optical images for bare (A), OTS- (B) and PYRENE-functionalized (C) wafers.

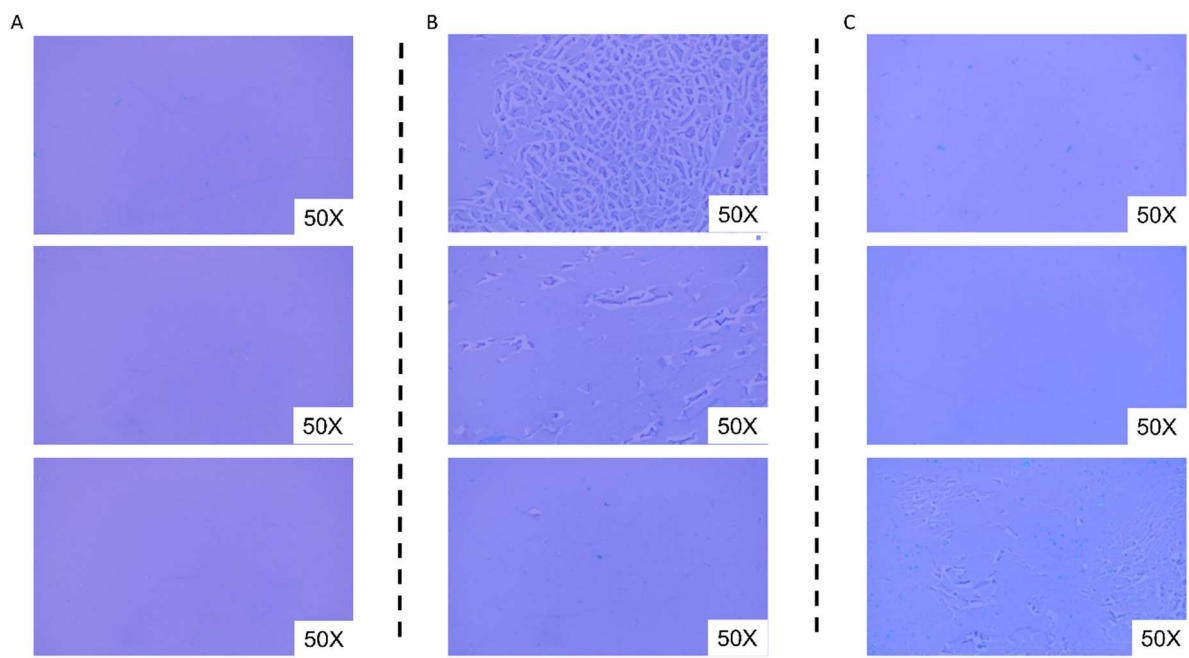

**Figure S12:** Zoomed in optical images (on three randomly chosen locations) for graphene on bare (A), OTS (B) and PYRENE (C) modified wafers on 4-inch wafer scale.

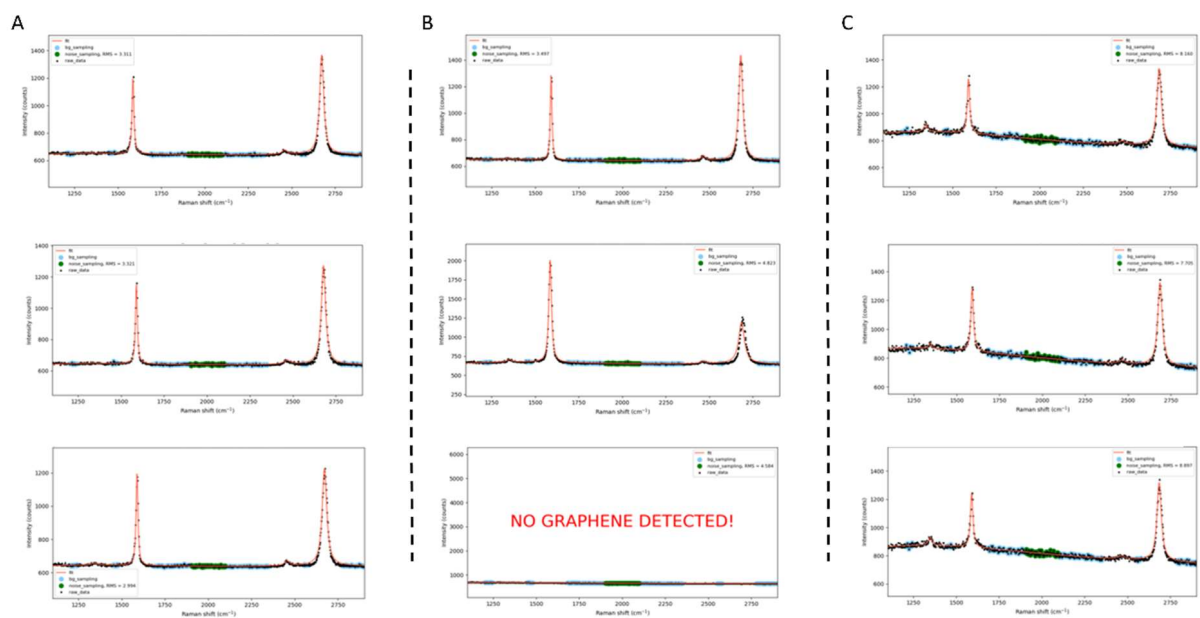

**Figure S13:** Raman spectra (on three randomly chosen locations) for graphene on BARE (A), OTS (B) and PYRENE (C) wafers on 4-inch wafer scale.

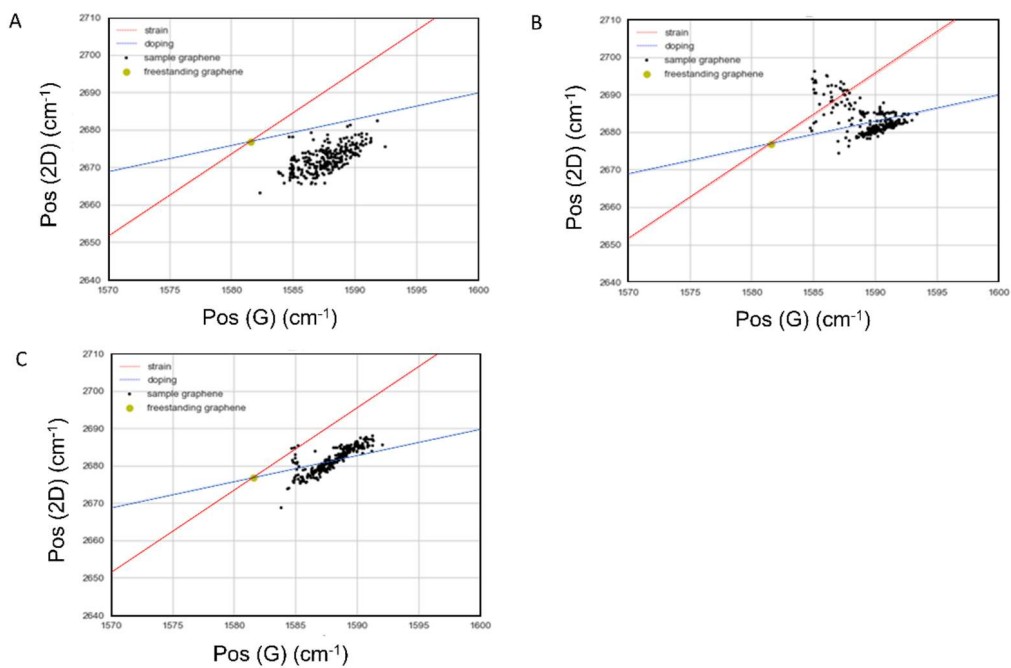

**Figure S14:** Raman spectroscopy G vs 2D scatterplot for graphene on BARE (A), OTS (B) and PYRENE (C) wafers on 4-inch wafer scale.

**Table S6:** Raman analysis of graphene on modified wafers on 4-inch wafer scale. Averages shown of spectra data that was suitable for fitting (indicated as #scans showing graphene). FWHM = full-width half maximum.

|  | BARE | | OTS | | PYRENE | |
|---|---|---|---|---|---|---|
| Parameter | Average | σ | Average | σ | Average | σ |
| Total Raman scans in measurement | 276 | N/A | 276 | N/A | 276 | N/A |
| Number of scans showing graphene | 274 | N/A | 261 | N/A | 274 | N/A |
| Percentage of scans with graphene | 99% | N/A | 94% | N/A | 99% | N/A |
| D position | 1343.36 | 5.5 | 1347.67 | 4.8 | 1347.44 | 5.1 |
| D intensity | 50.34 | 669 | 52.59 | 590 | 26.46 | 10 |
| D FWHM | 25.5 | 10 | 23.16 | 8 | 32.85 | 8 |
| G position | 1587.66 | 1.8 | 1589.66 | 1.9 | 1588.09 | 1.7 |
| G intensity | 622.78 | 138 | 675.97 | 523 | 490.3 | 44 |
| G FWHM | 11.75 | 1.6 | 12.4 | 2.6 | 16.54 | 1.2 |
| 2D position | 2672.21 | 3.3 | 2683.15 | 3.8 | 2681.73 | 3.3 |
| 2D intensity | 638.74 | 67 | 757.78 | 310 | 677.43 | 51 |
| 2D FWHM | 30.45 | 1.6 | 28.26 | 3.3 | 29.87 | 1.1 |
| G/D Ratio | 77.1 | 38 | 60.45 | 39 | 20.33 | 6 |
| G/2D Ratio | 0.98 | 0.2 | 0.93 | 0.5 | 0.73 | 0.1 |

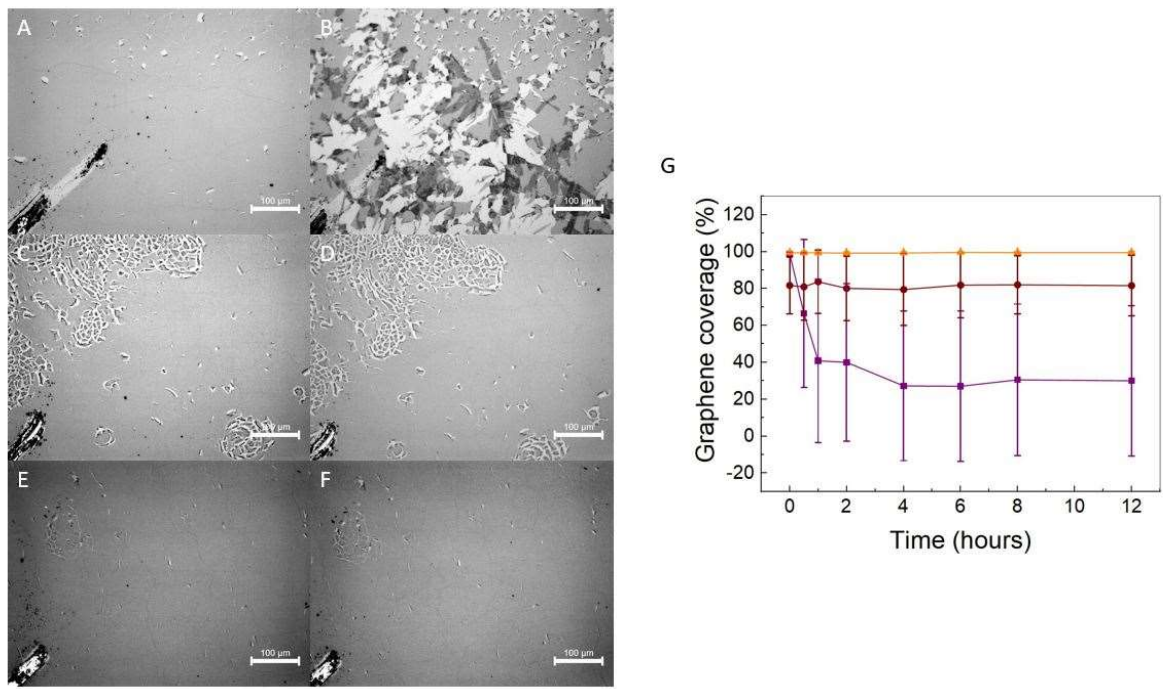

**Figure S15:** Visually enhanced optical images from graphene on modified wafers, immersed in glacial acetic acid at room temperature. BARE wafers before (A) and after 12 h (B), OTS wafers before (C) and after 12 h (D), and PYRENE wafers before (E) and after 12 h (F) immersion in acetic acid. G) Delamination vs. time for samples immersed in glacial acetic acid, for graphene on BARE, OTS and PYRENE wafers (purple, red, and orange respectively). Samples were optically inspected (three samples for each functionalization, four photos per sample) and delamination was determined as the percentage of area exposed in the center of the image (a circle with ⌀ = 150 μm).

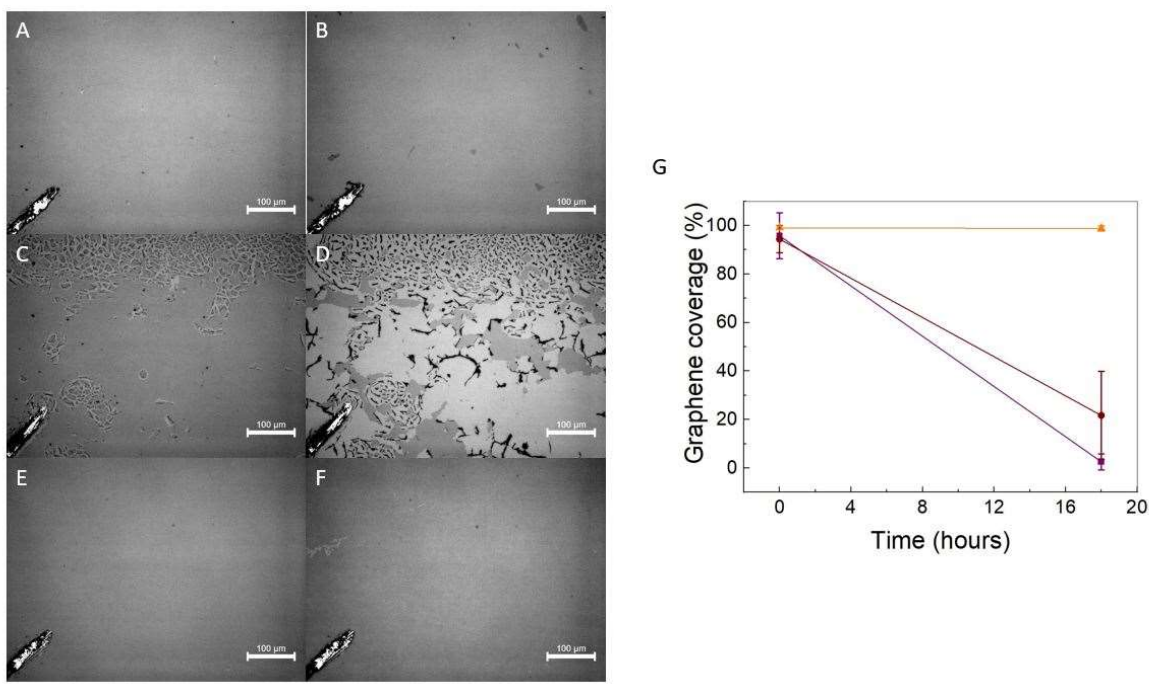

**Figure S16:** Visually enhanced optical images from graphene on modified wafers, immersed in N-methyl pyrrolidine (NMP) at room temperature. BARE wafers before (A) and after 12 h (B), OTS wafers before (C) and after 12 h (D), and PYRENE wafers before (E) and after 12 h (F) immersion in NMP. G) Delamination vs. time for samples immersed in NMP, for graphene on BARE, OTS and PYRENE wafers

(purple, red, and orange respectively). Samples were optically inspected (three samples for each functionalization, four photos per sample) and delamination was determined as the percentage of area exposed in the center of the image (a circle with ⌀ = 150 μm).

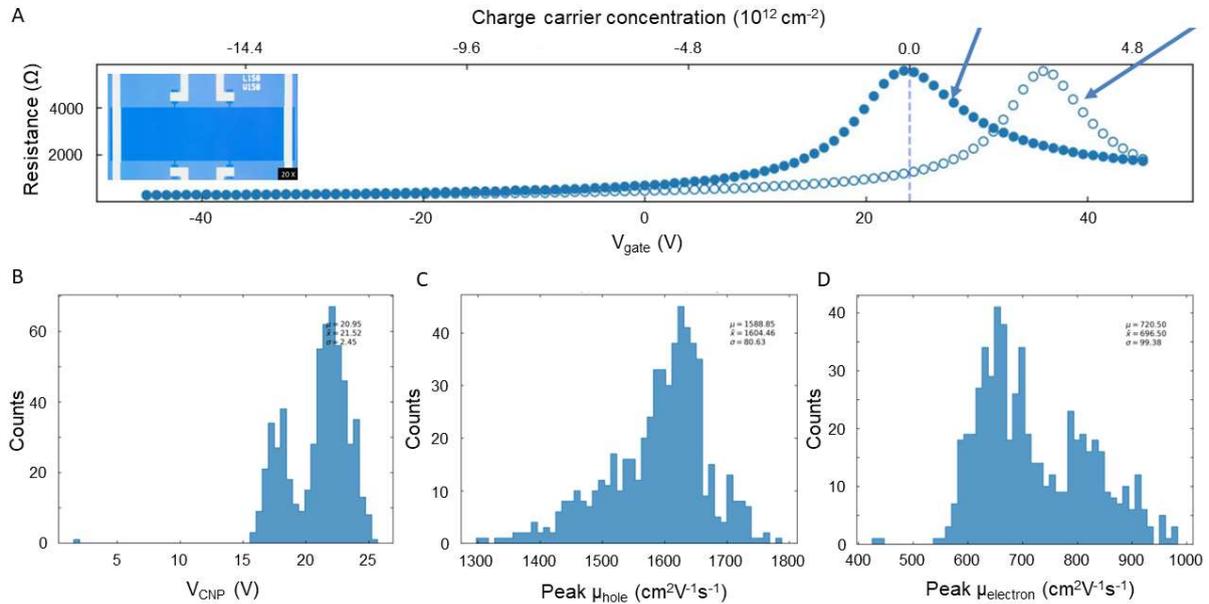

Figure S17: GFET characterization (4-terminal) of PYRENE wafer-scale device fabrication. A) Typical gate sweep for a GFET on a PYRENE wafer with 90 nm SiO$_2$ layer. Forward (backward) sweep in filled (empty) circles. B-D): $V_{CNP}$, $\mu_{hole}$, and $\mu_{electron}$ histograms for devices fabricated on a wafer with 90 nm SiO$_2$ thickness. Gate sweeps were performed at 10 V/s.

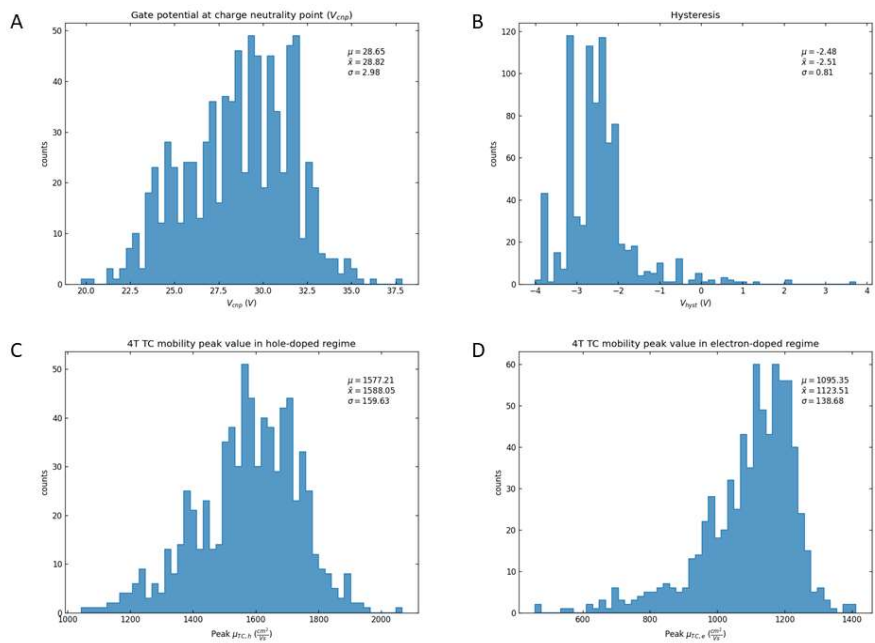

Figure S18: GFET characterization of BARE wafer-scale device fabrication. A-D): $V_{CNP}$, hysteresis, $\mu_{hole}$, and $\mu_{electron}$ histograms for devices fabricated on a wafer with 90 nm SiO$_2$ thickness. Gate sweeps were performed at 10 V/s.

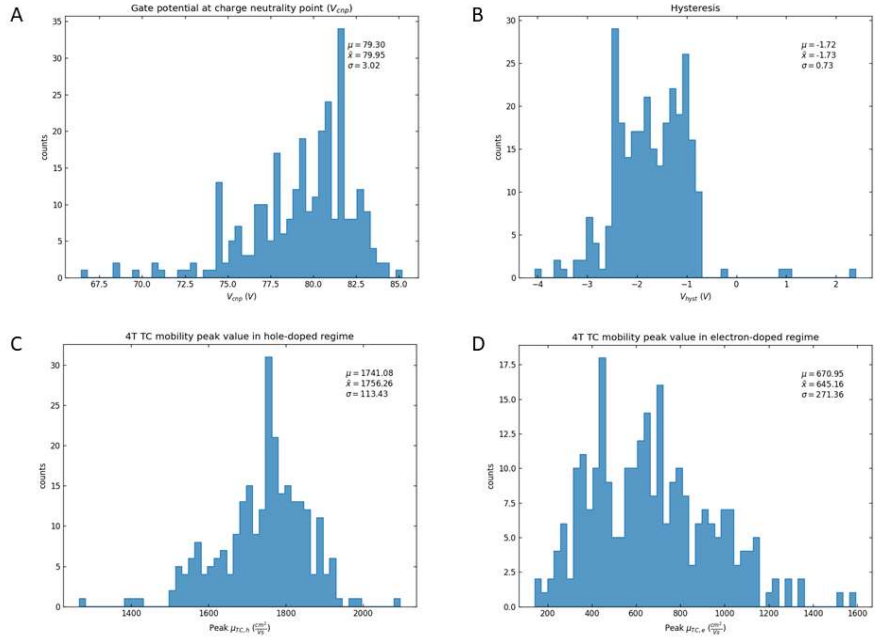

**Figure S19: GFET characterization of BARE wafer-scale device fabrication. A-D):** $V_{CNP}$, hysteresis, $\mu_{hole}$, and $\mu_{electron}$ **histograms for devices fabricated on a wafer with 285 nm SiO₂ thickness. Gate sweeps were performed at 10 V/s.**